\documentclass[aps,pre,preprint,groupedaddress]{revtex4}
\usepackage{graphicx}

\begin{document}


\title{Early stages of spreading and sintering}


\author{Scott T. Milner}
\email[]{stm9@psu.edu}
\homepage[]{https://sites.psu.edu/stm9research/}
\affiliation{Department of Chemical Engineering, The Pennsylvania State University, University Park, PA}


\date{\today}

\begin{abstract}
The early stages of sintering of highly viscous droplets are very similar 
to the early stages of a viscous droplet spreading on a solid substrate.
The flows in both problems are closely analogous to the displacements in a Hertzian elastic contact.
We exploit that analogy to provide both a scaling argument and a calculation
for the early growth of the contact radius $a$ with time, namely $a=(3 \pi \gamma R^2 t/(32 \eta))^{1/3}$.
(This result is complementary to the well-known Tanner law for spreading, $a \sim t^{1/10}$,
which holds in the regime of low contact angles.)
For viscoelastic fluids, the linear scaling of $a^3$ with time
is replaced by the general result that $a^3(t)$ is proportional to the creep compliance $J(t)$.
\end{abstract}


\maketitle


\section{Introduction.}
Sintering occurs when two liquid droplets are drawn together by capillary forces, 
i.e., by the reduction in interfacial energy that accompanies the merging of the two droplets.  
Sintering is a key element in several commercially important manufacturing processes,
for example rotational molding \citep{rotomolding} and the formation of continuous latex films \citep{Lovell}.

In the simplest case, we consider two equal-sized spherical droplets of Newtonian fluid.
For sintering of polymer particles only slightly above the glass transition temperature, 
the viscosity will be very large, and the sintering process very slow.
Thus the effect of lubrication forces from the expulsion of the intervening fluid,
which we assume to be of ordinary viscosity, will be negligible here.
This is in contrast to the usual case of the coalescence of two fluid droplets
suspended in a second immiscible fluid, in which the viscosities are comparable.

The dynamics of the early stages of sintering are very similar 
to a viscous droplet spreading on a flat solid substrate,
whether in air or suspended in fluid.
One may regard the early stages of sintering of two droplets
as one droplet spreading on the nearly-flat surface presented by the other droplet.
The flow in the vicinity of the contact is predominately in the normal direction,
as the droplets flatten to quickly increase their mutual contact area.

In each case, the driving force is the reduction in interfacial energy;
for the spreading droplet, the relevant combination of interfacial tensions at early times
is $\Delta \gamma = \gamma_{LV} + \gamma_{SV} - \gamma_{LS}$
(where $L, V, S$ denote respectively liquid, vapor, and substrate).
For the sintering droplets, the driving force is simply $\gamma_{LV}$.

There are differences between the two situations; 
evidently, the boundary conditions on the contact plane are different,
in that the spreading droplet must have zero velocity on the substrate,
whereas for sintering droplets the velocity in the plane of contact can have a radial component.

As well, for spreading droplets the changing geometry and competing interfacial tensions 
can lead to a finite contact angle, 
obeying the Young equation for mechanical equilibrium of the contact line,
\begin{equation}
0 = \gamma_{SV} - \gamma_{SL} - \gamma_{LV} \cos \theta
\label{Young_eqn}
\end{equation}
Still, whether or not the equilibrium contact angle is finite,
the initial dynamics of the contact, when the dynamic contact angle is close to $\pi$,
should correspond well with the sintering of two identical droplets,
with $\Delta \gamma$ playing the role of $\gamma_{LV}$.

For spreading or sintering of polymeric fluids,
a more realistic treatment should represent the fluids as viscoelastic,
with a time-dependent creep compliance.
The simplest representation of such a viscoelastic fluid 
would be in terms of a Maxwell model 
with a rubberlike modulus and a single relaxation time.
For sintering of polymer particles slightly above $T_g$, 
we may expect time-temperature superposition to hold.
That is, the viscoelastic relaxation time, indeed all rheological times,
would have the same dependence on $T-T_g$.
Thus the creep response would have the same functional form at all temperatures,
with a characteristic time that depended on $T-T_g$.

There is a strong analogy between the early stages of spreading or sintering and Hertzian contacts,
such as appear in the theory of Johnson, Kendall, and Roberts \cite*{JOHNSON:1971p10405} 
for incompressible but soft elastic spheres in adhesive contact.
In the JKR theory, the spherical particles are solid, not liquid,
so the deformation of the particles is time-independent and to be calculated.
The recipe for the JKR calculation is to suppose that a pattern of forces 
is imposed on each particle over the flat circular contact area, 
such that 1) the displacement field over the contact region is consistent with a flat contact,
and 2) the total force over the contact equals the adhesive force.
The adhesive force is determined by a virtual work argument, 
i.e., by considering the variation of the adhesive energy 
(arising from the variation in the contact area, times some adhesive energy per area)
with respect to the separation between the particle centers.

In the spreading problem for a Newtonian fluid, 
the contact between the droplets is likewise a flat circular area,
and a pattern of forces likewise acts on each droplet only over that area.
Of course, the deformation is time-dependent, 
well approximated by creeping flow (zero Reynolds number).
So we compute the flow field, not the displacement field,
and the JKR recipe is replaced by the requirements that 
1) the deformation field over the contact region
is consistent with a flat circular contact growing at some rate,
and 2) the total force over the contact equals the capillary force.
(The capillary force, essentially identical to the adhesive force in the JKR model,
is determined by a virtual work argument.)

In this paper we are concerned exclusively with the early stages of spreading;
the more familiar regime for spreading of completely wetting fluids,
in which the contact radius $a$ far exceeds the initial droplet radius $R$,
is described by the Tanner law \citep{TANNER:1979p8718},
\begin{equation}
\frac{a(t)}{R} \sim \left( \frac{\gamma R t}{\eta} \right)^{1/10}.
\end{equation}
in which $\eta$ is the fluid viscosity and $\gamma$ the interfacial tension.

This paper is organized as follows.  
In Section \ref{static_scaling}, we reprise scaling arguments for a Hertzian elastic contact,
and in Section \ref{Hertz_calc}, the detailed calculation of the displacements and normal stress fields.
In Section \ref{spreading_scaling}, we present analogous scaling arguments 
for the spreading at early times viscous droplet, 
and in Section \ref{spreading_calc} the detailed calculation of the velocity and normal stress fields.
In Section \ref{sintering_scaling}, we argue that the differences between sintering and spreading
lead to corrections to scaling at early times, thus our calculations should apply to both situations.
Section \ref{spreading_expt} discusses recent experimental results for viscous droplet spreading at early times.
We extend our treatment to general viscoelastic fluids in Section \ref{viscoelastic}.
In Section \ref{previous_work} we compare our results 
to previous analytical, numerical, and experimental findings on Newtonian and viscoelastic sintering.  
(We reverse the conventional order, discussing prior results after presenting our approach, 
because in the present case this simplifies the discussion of a somewhat confusing and contradictory literature.)
General conclusions and suggestions for future experiments are presented in the final Section.

\section{\label{static_scaling}Scaling arguments.}
The mathematics of the Hertzian contact \citep{Hertz} are a bit involved,
so it is useful to begin with a scaling argument that anticipates some of the results.
It turns out that the geometry of the Hertzian contact (see Fig. \ref{hertz_contact}), 
with its deformation confined primarily to a small spherical cap,
leads to nontrivial scaling for the relation between the force (or interfacial tension)
and the interparticle separation (or contact area).
For the Hertz problem, we suppose that some body force $F$
acts on each particle to push it against the other.
Later, we will make a scaling argument as to the magnitude of $F$
arising from adhesive energy.

The following scaling assumptions may be defended as plausible a priori,
and are consistent with the Hertz solution:
\begin{equation}
a^2 \sim R h
\end{equation}
The geometry of the contact area, a circle of radius $a$, is that of a spherical cap; 
here $R$ is the radius of the particles, 
and $h$ is the width of the gap between the particles,
 i.e., $2R-h$ is the separation between particle centers.

\begin{equation}
F \sim \sigma a^2
\end{equation}
The total force acting between the particles 
scales as the typical value of the normal stress $\sigma$ times the contact area.

\begin{equation}
\sigma \sim \mu \epsilon
\end{equation}
The stress scales as the shear modulus $\mu$ times the typical strain $\epsilon$.

\begin{equation}
\epsilon \sim h/a
\end{equation}
The strain is the gradient of the displacement,
which varies from a value of order $h$ at the contact center,
to a value of order zero at the contact edge, a distance $a$ away.
Thus displacements scale as $h$ in a Hertzian contact, and gradients scale as $1/a$.
The displacement field is localized to the contact region, 
and is negligibly small over the main volume of the particle. 

\begin{equation}
U \sim F h
\label{U_elastic}
\end{equation}
The stored elastic energy is the work done by the force acting over the displacement.

Combining these scaling assumptions leads to the following results:
\begin{eqnarray}
a &\sim& F^{1/3} R^{1/3} \mu^{-1/3} \nonumber \\
h &\sim& F^{2/3} R^{-1/3} \mu^{-2/3} \nonumber \\
U& \sim& h^{5/2} R^{1/2} \mu
\end{eqnarray}
The above scaling relations, which encode only the geometry of the spherical cap
and locally linear stress-strain relations,
result in nonlinear relations between the contact radius $a$ (or area $a^2$) and force $F$,
or between the displacement $h$ and the force $F$,
or between the elastic energy $U$ and the displacement $h$.
The Hertzian ``spring'' is nonlinear, because the more you push into it,
the greater volume of elastic material is engaged in pushing back.

Note that it is equivalent to assume instead of Eqn.\ (\ref{U_elastic}) above,
that the stored elastic energy scales as the volume $a^3$, 
times the strain energy density $\mu \epsilon^2$:
\begin{equation}
U \sim a^3 \mu \epsilon^2
\end{equation}
The volume over which elastic energy is stored is $a^3$ not $a^2 h$,
because the gradients are all of order $1/a$ --- 
the strain field extends vertically into droplet a distance of order $a$, not $h$.

Now consider the adhesive force, which arises from the adhesive energy of contact $W$,
which scales as
\begin{equation}
W \sim -\gamma a^2
\end{equation}
in which $\gamma$ is the adhesive energy per unit area.
We can rewrite this in terms of $h$ using the spherical cap scaling above,
whereupon $W \sim -\gamma h R$.  
The corresponding adhesive force, which by virtual work arguments
is the derivative $-dW/dh$, is evidently constant, independent of $h$.
So adhesive forces lead to a constant total load.

\section{\label{Hertz_calc}Details of Hertz solution.}
Here we reprise the Hertz solution, 
specialized to the case of incompressible isotropic solid spheres of equal radius.
This development follows Landau and Lifshitz \cite{LL_elasticity}.  
We present it here because 
1) the incompressible case for spheres of equal size is simpler than the general case,
and 2) we shall use these results to develop an analogous calculation 
for sintering of incompressible Newtonian fluid droplets of equal radius.

Incompressible isotropic solids satisfy an equilibrium equation,
\begin{equation}
\nabla \cdot \sigma + f = 0
\end{equation}
For small displacements, the stress tensor $\sigma_{ij}$ in terms of the displacement field $u_j$
takes the form
\begin{equation}
\sigma_{ij} = \mu (\nabla_i u_j + \nabla_j u_i) - p \delta_{ij}
\end{equation}
The pressure $p$ is determined by the incompressibility condition.

This leads to equations
\begin{eqnarray}
&\mu \nabla^2 u + f = \nabla p & \nonumber \\
&\nabla \cdot u = 0 &
\end{eqnarray}
The Green function for this system of equations is the Oseen tensor
\begin{equation}
J_{ij}(x) = \frac{1}{8 \pi \mu} \left( \frac{\delta_{ij}}{|x|} + \frac{x_i x_j}{|x|^3} \right)
\end{equation}

In terms of the Oseen tensor, we may write the solution $u(x)=F \cdot J(x)$ 
for a point force $f(x) = F \delta(x)$ applied at the origin.
The corresponding pressure field is $p(x) = F \cdot x/(4 \pi |x|^3)$.

More generally, if forces are applied over some region, 
the displacement field will be a convolution of the Oseen tensor with the force density,
\begin{equation}
u(x) = \int dx' \, J(x-x') \cdot f(x')
\label{Green_eqn}
\end{equation}

\begin{figure}
\begin{center}
\includegraphics[scale=1.0]{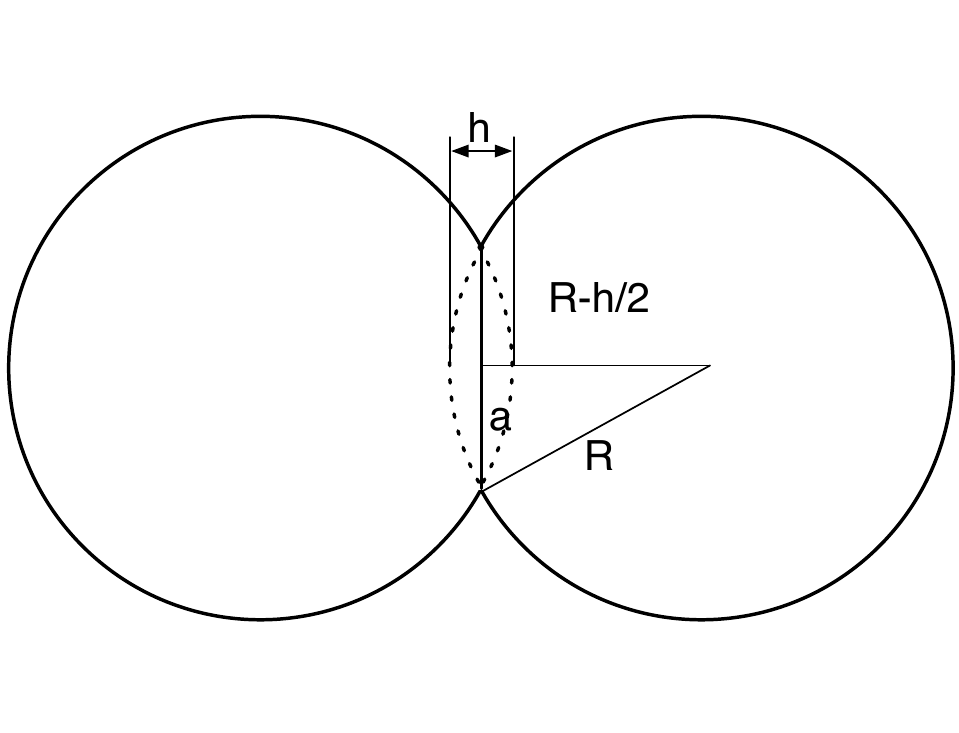}
\caption{\label{hertz_contact}Geometry of the Hertzian contact of radius $a$,
between two equal-sized spheres of undeformed radius $R$.}
\end{center}
\end{figure}

In the Hertz problem, forces are applied only on the small contact area (radius $a$)
of large spherical particle (radius $R>>a$).
By rotational symmetry about the line between centers,
the force density can only depend on the in-plane radial distance $r$ 
from the central point of contact.
The problem is to find the force density $F(r)$, zero outside the contact area, 
that gives a displacement field consistent with a flat contact,
and has a total integral over the contact area equal to the applied load.  (See Fig.\ \ref{hertz_contact}).

Over the area of contact, the applied force can only have a normal component.
Any radial component would have the same magnitude for the two spheres
by reflection symmetry, and so could not be a ``reaction pair'', equal and opposite.
With the applied force in the normal ($z$) direction only,
the form of the Oseen tensor implies
that only the $z$ component of the displacement vector $u(x)$ 
is nonzero along the plane of the applied force. 

At lowest order in $a/R$, we can compute the strain field in response to the applied normal force
in the approximation that the undeformed interface is strictly flat.
Thus we have a normal force applied on a finite region of the boundary of an elastic half-space.
The Oseen tensor provides the Green function for the closely related problem
of a force applied in the $z$ direction on a portion of the x-y plane inside an infinite medium.
There, we see that the displacements on each side of the plane are symmetrically disposed,
such that the normal displacements are equal and the transverse displacements opposite in sign.
Thus, the normal stresses are equal and opposite in the two half-spaces that together constitute the infinite medium.

We may then give the solution for normal forces applied to the boundary of a single half-space
by employing the infinite medium solution, 
and regarding half the force as acting on one half-space and half on the other. 
Thus Eqn.\ (\ref{Green_eqn}) holds as a boundary integral, 
but with an additional factor of two:
\begin{equation}
u(x) = 2 \int d^2x' \, J(x-x') \cdot f(x')
\label{Green_eqn2}
\end{equation}
On the boundary itself, the displacement is only in the normal direction:
\begin{equation}
u_r(r,z=0) = \frac{1}{4 \pi \mu} \int_{r'<a} d^2 r' \frac{f_z(r')}{|r-r'|}
\label{displacement}
\end{equation}

The displacement in the region of the contact area 
must be such that it flattens a spherical cap into a planar contact.
The spherical cap geometry implies that 
\begin{equation}
u_z(r,z=0) = \frac{h}{2} \left( 1 - \frac{r^2}{Rh} \right)
\label{displacement_quadratic}
\end{equation}

So we seek a function $f_z(r')$ that leads to a quadratic for $u_z(r)$.
At this point, following LL:  
1) we recognize that the right-hand side has the form of an integral
for the electrostatic potential in three dimensions 
of a thin disc-shaped charge distribution of thickness $f_z(r')$, evaluated in the plane of the disk ($z=0$);
2) we recall that the potential of an ellipsoidal charge distribution
of constant charge density is a quadratic function of the coordinates from the center.

The potential inside such an ellipsoid of constant charge density $\rho$
is given by a result from ellipsoidal coordinates,
\begin{equation}
\phi(x,y,z) = \pi \rho a b c \int_0^\infty \left[ 1 - \frac{x^2}{a^2+\xi} - \frac{y^2}{b^2+\xi} 
- \frac{z^2}{c^2+\xi} \right] \frac{d \xi}{\sqrt{(a^2+\xi)(b^2+\xi)(c^2+\xi)}}
\end{equation}

For present purposes, we take $a=b$ (cylindrical symmetry) and the limit of small $c$ as well as $z=0$,
which gives the potential inside a highly flattened ellipsoid of revolution on the central plane, as
\begin{equation}
\phi(r) = \pi \rho a^2 c \int_0^\infty \left[ 1 - \frac{r^2}{a^2+\xi} 
 \right] \frac{d \xi}{(a^2+\xi)\sqrt{\xi}}
\end{equation}

We see that this potential is indeed a quadratic function of $r$.
In fact, on substituting $\xi = s^2$ the integrals are elementary, with the result
\begin{equation}
\phi(r) = \pi^2 \rho a c \left( 1 - \frac{r^2}{2a^2} \right)
\end{equation}

Of course, this potential can also be written as an electrostatic integral,
\begin{equation}
\phi(x,y,z) = \int \frac{\rho dx' dy' dz'}{|r-r'|}
\end{equation}
In the limit of a squashed ellipsoid of revolution, 
and computing the potential on the central plane, this can be recast as
\begin{equation}
\phi(r) = 2 \rho c \int \frac{d^2 r'}{|r-r'|} \sqrt{1-\frac{r'^2}{a^2}}
\end{equation}
In the above, the numerator results from the local ``thickness'' of the charged disk.

So now we enforce a choice of coefficients 
such that $u_z(r)$ and $\phi(r)$ are identical.
Comparing the computed results for each,
this implies $h/2 = \pi^2 \rho a c$ and $2a^2 = Rh$
(which is consistent with the scaling arguments).

Comparing the integral expressions 
in terms of the corresponding force and charge densities, we must have
\begin{equation}
f(r) =  \frac{4\mu h}{\pi a} \sqrt{1-\frac{r^2}{a^2}}
\label{force_density}
\end{equation}

With this result, we can compute the integral over the contact 
of the force areal density, to find the total force:
\begin{equation}
F = 2 \pi \int_0^a r dr f(r) = \frac{16}{3} \frac{a^3 \mu}{R}
\end{equation}
The above result is consistent with the scaling arguments.

We may write the adhesive energy (with coefficients) as
\begin{equation}
W = -\pi \gamma a^2 = (\pi/2) \gamma Rh
\end{equation}
so that the adhesive force $F_{adh} = -dW/dh$ is simply $F_{adh} = (\pi/2) \gamma R$.
Then, equating the adhesive force to that required to deform the particle, 
we find after a bit of algebra
\begin{equation}
a = \left( \frac{3 \pi \gamma R^2}{32 \mu} \right)^{1/3}
\label{spreading_result}
\end{equation}

\begin{figure}
\begin{center}
\includegraphics[scale=0.7]{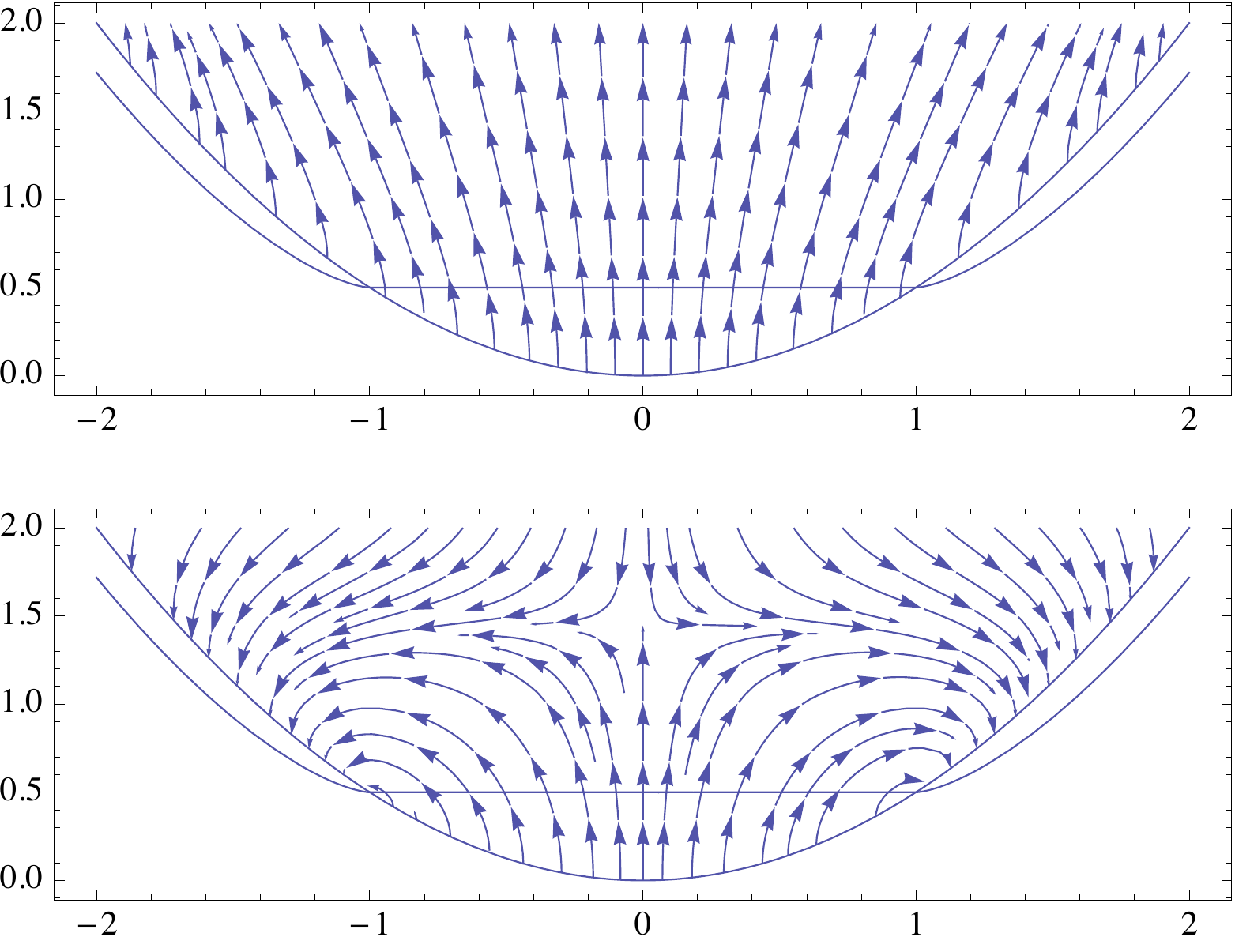}
\caption{\label{u_fig}Streamline plot of deformation field $u(r)$ for Hertzian contact, 
in reference frame with fixed (a) particle center; (b) contact plane.  Solid curves,
sphere surface before and after deformation.}
\end{center}
\end{figure}

The displacement fields for the Hertzian contact can be calculated from Eqn.\ (\ref{Green_eqn2}); 
the results are displayed as a streamline plot in Fig.\ \ref{u_fig}.
In the figure, the displacements are indicated 
relative to the (parabolic approximation of the) undeformed sphere,
for reference frames in which either (a) the center of the particle is fixed,
or (b) the contact plane is fixed.

\section{\label{spreading_scaling}Scaling for spreading.}
Before we begin the detailed calculation of Stokes flow in the early stages 
of a single droplet spreading on a planar substrate,
we again present a physical scaling argument that anticipates the basic results.
We shall argue in Section \ref{sintering_scaling} below
that the same scaling considerations apply to the early stages of sintering.
(In Section \ref{sintering_scaling}, we present scaling arguments to show 
that the differences between sintering and spreading flows,
associated with the additional capillary forces on the neck due to curvature at the joint between particles,
are a correction to scaling at early times.)

Evidently, dimensional analysis alone does not get us very far in this problem;
all we can say is that the parameters $\eta$, $R$, and $\gamma$ determine a characteristic time,
\begin{equation}
\tau = \eta R/\gamma
\end{equation}
We expect the contact radius $a(t)$ therefore to grow as
\begin{equation}
a(t)/R \sim (t/\tau)^\alpha
\label{dim_analysis}
\end{equation}
but dimensional analysis alone gives no indication as to the value of $\alpha$.
Even the Tanner law takes this form, with $\alpha=1/10$.

The scaling description of the early stages of droplet spreading
is quite analogous to our treatment of the Hertzian contact.
As two Newtonian droplets combine under the action of capillary forces,
the spherical cap geometry again applies, 
so that the contact radius $a$, the particle displacement beyond contact $h$, 
and the particle radius $R$, are related by
\begin{equation}
a^2 \sim hR
\end{equation}

The origin of the attractive force $F$ is again 
the reduction of interfacial energy $W$ as the droplets merge;
we again have 
\begin{equation}
W \sim -\gamma a^2
\end{equation}
and by virtual work arguments 
\begin{equation}
F \sim -\partial W/\partial h \sim \gamma R
\end{equation}
Thus the force is again constant with respect to particle displacement.

The rate of work done by this force as the particles approach is
\begin{equation}
\dot W \sim F \dot h
\end{equation}
(which may also be cast as $\dot W \sim \gamma d a^2/dt$).
This must balance the rate of viscous dissipation within the particle,
\begin{equation}
\dot W \sim \eta \dot \epsilon^2 \Omega
\end{equation}
in which $\dot \epsilon$ is the typical strain rate 
and $\Omega$ is the volume over which the dissipation occurs.

The gradients of the Stokes flow are of order $1/a$ and the volume $\Omega$ of order $a^3$
(as for the Hertzian contact problem),
corresponding to the spatial dependence of fields satisfying a Laplace equation in three dimensions.
The strain rate is then $1/a$ times a typical velocity of order $\dot h$; hence
\begin{equation}
\dot \epsilon \sim \dot h/a
\end{equation}

Putting these results together, we find
\begin{equation}
a(t) \sim \left( \frac{ F R t}{\eta} \right) ^{1/3}
\end{equation}
which evidently has the same power-law dependence on $F$ and $R$ 
as did the Hertzian contact result.
Using the scaling for the adhesive force, we have
\begin{equation}
\frac{a(t)}{R} \sim \left( \frac{\gamma t}{\eta R} \right)^{1/3}
\end{equation}
which is of the anticipated form Eqn.\ (\ref{dim_analysis}), with $\alpha=1/3$.

Note that this scaling result does not agree with that of Frenkel \cite{Frenkel}, 
who long ago presented a scaling argument 
for the growth of the contact radius for sintering droplets.
He obtained
\begin{equation}
\frac{a(t)}{R} \sim \left( \frac{\gamma t}{\eta R} \right)^{1/2}
\end{equation}
But his result is based on scaling assumptions that appear to be faulty; namely,
\begin{eqnarray}
& \dot \epsilon \sim \dot h/R &  \nonumber \\
& \Omega \sim R^3 &
\end{eqnarray}
In other words, the characteristic length scale for gradients 
was incorrectly taken by Frenkel to be $R$, not $a$.
It is clear, however, that the flow fields associated with the deformation of the small contact
do not penetrate appreciably over the entire droplet volume.
Another way to reach this errant result, as we shall see below, 
is to assume $\Omega \sim a^2 h$ and $\dot \epsilon \sim \dot h/h$;
that is, that the volume over which dissipation occurs is the ``contact lens'',
with a volume of order $a^2 h$,
and that the flow gradients are of order $1/h$.

\section{\label{spreading_calc}Spreading of Newtonian droplets.}
Now we present detailed calculations for early-time spreading Newtonian droplets.
We observe that the force-balance equations for incompressible Stokes flow
take exactly the same form as those for deformation of an incompressible solid.
The displacement field $u$ becomes the flow velocity $v$,
the shear modulus $\mu$ becomes the shear viscosity $\eta$.

However, the condition on the form of the displacement field is changed.
Instead of requiring that the displacement field $u_z$ 
is such that it flattens a spherical cap,
we impose the condition that the velocity field $v_z$ over the contact area 
must be constant, equal to $\dot h/2$.
(With this choice of boundary condition,
we are working in a frame in which the position of the droplet center is fixed,
and the substrate moves vertically up against the droplet as it flattens.
In the lab frame, we add $-\dot h/2$ to $v_z$ everywhere,
so that $v_z=0$ on the stationary substrate, and the droplet center moves downwards.) 

So we must construct another charge distribution, 
this time leading to a constant potential inside the contact area.
Looking at the functional form for $u_z$ and recalling that $h$ scales as $a^2$,
we see that taking a derivative $a^2 \partial/\partial a^2$ 
of our result Eqn.\ (\ref{displacement}, \ref{displacement_quadratic}) $u_z(r)$
would give a constant result $h/2$, independent of $r$.

So we take this same derivative of the potential integral,
to find the corresponding charge density.
Note that $\rho c$ scales as $h/a$ or equivalently $a/R$,
so that a derivative $a^2 \partial/\partial a^2$ applied to Eqn.\ (\ref{displacement})
acts on the function $f(r)$ [given by Eqn.\ (\ref{force_density})]
in two places, the overall prefactor and under the radical.

Acting on the prefactor scaling as $a$, the derivative gives a factor of 1/2;
acting on the radical $(1-r^2/a^2)^{1/2}$, we obtain $(1/2)(r^2/a^2)(1-r^2/a^2)^{-1/2}$.
After combining these two terms, we find a correspondence between a potential
\begin{equation}
\phi(r) = \pi^2 \rho a c = \rho c \int \frac{d^2r'}{|r-r'|} \frac{1}{\sqrt{1-r'^2/a^2}}
\end{equation}
and a velocity field
\begin{equation}
u_z(r) = \dot h/2 = \frac{1}{4 \pi \eta} \int_{r'<a} d^2r' \frac{f_z(r')}{|r-r'|}
\label{velocity}
\end{equation}

If we replace $h \rightarrow \dot h$ and $u_z \rightarrow v_z$,
the same correspondence as before ($\pi^2 \rho a c = h/2$)
leads to the identification of the force areal density 
for the case of viscous sintering as
\begin{equation}
f_z(r) = \frac{2 \eta \dot h}{\pi a} \frac{1}{\sqrt{1-r^2/a^2}}
\label{force_density2}
\end{equation}

The corresponding total force is again obtained 
by integrating $f_z$ over the contact area, with the result
\begin{equation}
F = 4 \eta \dot h a
\end{equation}

We have argued above in the discussion on adhesive contact
that the force resulting from reduction in interfacial energy is constant,
independent of the displacement of the two particles towards each other.
This is equally true for the case of spreading.
We may take the same expression for the reduction in interfacial energy
resulting the reduction in interfacial area, which occurs as the droplets merge.  

Using $hR=2a^2$ to replace $\dot h$ in the expression for the force,
we find after a bit of arithmetic that $a^3$ grows linearly in time,
\begin{equation}
a(t) = \left( \frac{3 \pi \gamma R^2 t}{32 \eta} \right)^{1/3}
\label{contact_vs_time}
\end{equation}
so that $h$ (and the contact area) grows as $t^{2/3}$.

Note that this analysis neglects any stresses arising from the lubrication flow
of the surrounding fluid moving out of the gap between the spreading droplet and the substrate,
which should be valid if the droplet is much more viscous than the surrounding fluid.

\begin{figure}
\begin{center}
\includegraphics[scale=0.7]{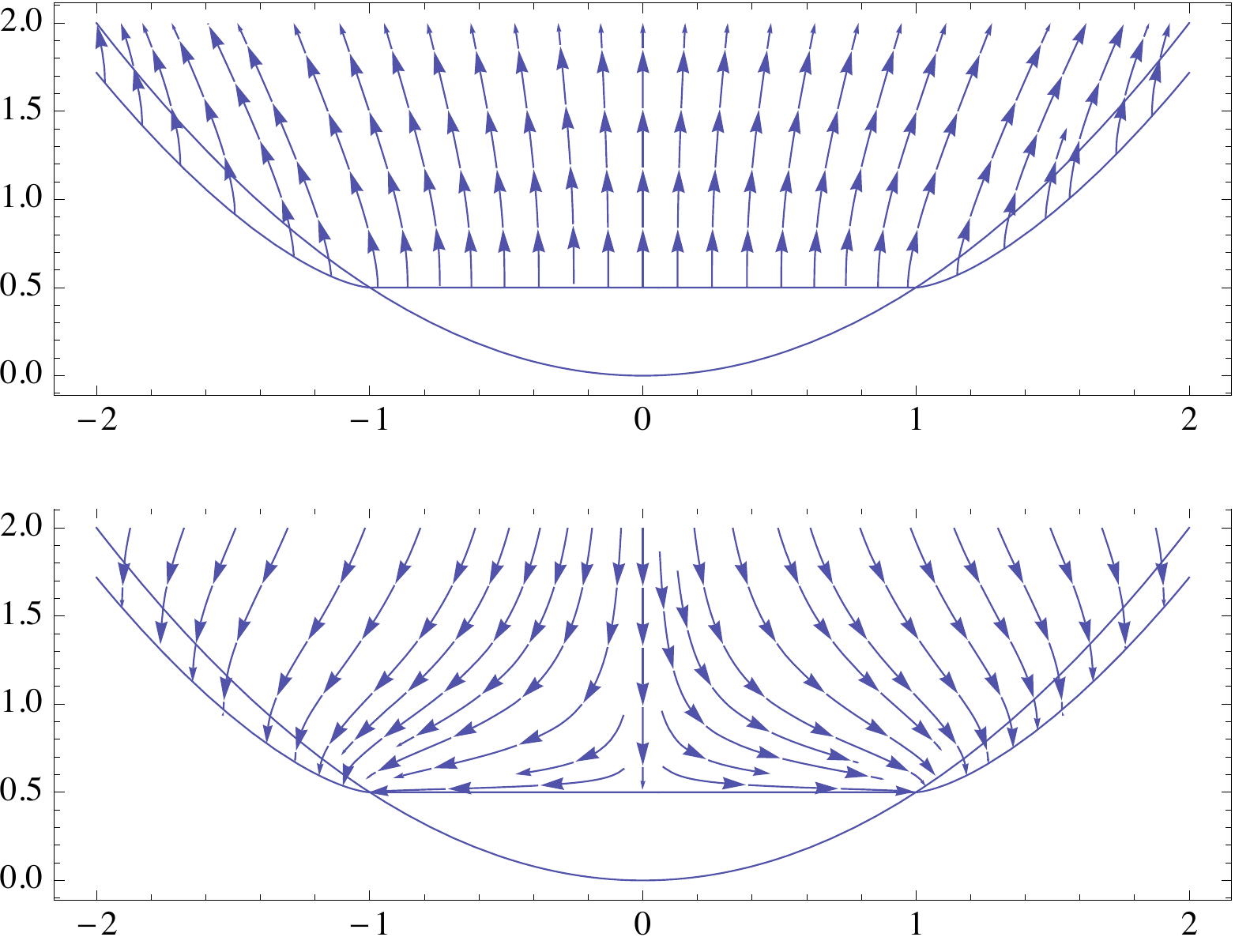}
\caption{\label{v_fig}Streamline plot of velocity field $v(r)$ for early stage sintering or spreading,
in reference frame with fixed (a) particle center; (b) contact plane.  
Solid curve indicates surface of deformed sphere.}
\end{center}
\end{figure}

The velocity fields for the spreading calculation are given from Eqn.\ (\ref{velocity}); 
the results are displayed as a streamline plot in Fig.\ \ref{v_fig}.
The flows in Fig.\ \ref{v_fig} are indicated relative to the deformed sphere,
again for reference frames in which either (a) the center of the particle is fixed,
or (b) the contact plane is fixed.

\begin{figure}
\begin{center}
\includegraphics[scale=0.6]{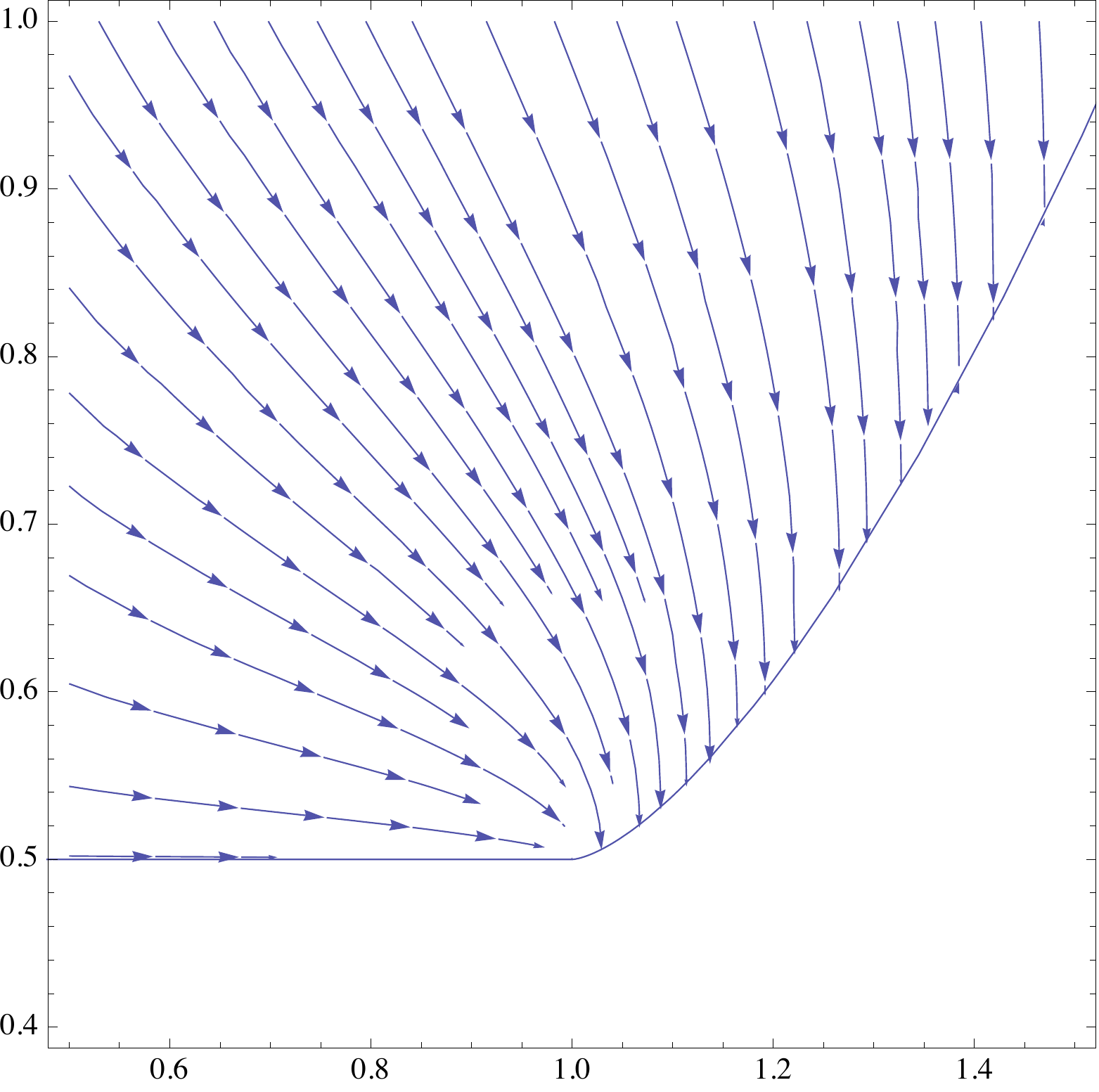}
\caption{\label{closeup}Closeup of streamline plot of Fig.\ \ref{v_fig} near contact rim [point (1,1/2) in figure].}
\end{center}
\end{figure}

A closeup of the flow field near the contact point is shown in Fig.\ \ref{closeup},
in the reference frame in which the contact area is fixed in space
(for reference, also shown is the undeformed shape of the particle, 
as the smooth curve extending below the flow field to the origin).
It is evident that the flow is predominately ``down from above'' on the outside (right side)
of the contact point [at (1,1/2) in the figure], which tends to close the gap 
and thereby increase the contact area.

\section{\label{sintering_scaling}Sintering versus spreading.}
We claim that the calculations of the previous section apply equally 
to the early stages of droplet spreading on a planar interface,
and to the early stages of sintering of two identical spherical droplets.
Indeed, sintering may be regarded approximately as droplets spreading on each other.
The relevant interfacial tension for sintering is $\gamma_{LV}$,
the energy of the interface between the liquid droplet (L) and surrounding vapor (V).
(Two such interfaces are lost when droplets merge,
but there is dissipative flow in each droplet, so each droplet ``owns'' one of the interfaces.)
For spreading, the relevant tension is $\gamma_{LV} + \gamma_{SV} - \gamma_{LS}$,
where S denotes the substrate.

Of course, potentially important differences between the early stages of spreading and sintering
may arise from the different boundary conditions in the two situations.
In the case of spreading, the velocity field must vanish on the substrate,
and the surface of the droplet can meet the substrate at a finite contact angle.

In contrast, for sintering the velocity on the contact plane is not obliged to vanish,
but can have an outward radial component.
In fact, there is a driving force pulling radially outwards on the contact rim,
arising from capillary pressure associated with the curvature of the ``neck'' between the droplets
in the direction transverse to the contact plane,
tending to smooth out the cusp at the contact rim evident in the sketch of Fig.\ \ref{hertz_contact}.
Physically, the surface energy of sintering droplets is reduced as this curvature decreases.
To reduce this curvature, material must flow towards the contact line to ``fill in'' the cusp.
In this section, we construct a scaling argument for the rate of growth of the neck as a result of this secondary flow.
We show that in the early stages of sintering, this ``cusp smoothing'' mechanism 
is a correction to the ``droplet flattening flow'' we have computed.

Scaling arguments have been given previously for the growth rate 
of the contact area by capillary pressure at the neck.
Indeed, Skorokhod \cite{Skorokhod:1995p10580} later identified 
the problems with the Frenkel scaling assumptions
and attempted a revised justification for the Frenkel scaling
(which are often roughly consistent with published observations), 
by constructing a scaling argument for the growth rate by this mechanism.

Skorokhod argued (incorrectly, we shall assert)
that the relevant dissipation volume for a flow associated with neck formation
was $\Omega \sim a^2 h$, 
and that the characteristic strain rate was $\dot \epsilon \sim \dot h/h$.
The driving force claimed for neck formation 
is once again the reduction in interfacial area,
which leads to a proposed balance
\begin{equation}
\gamma d a^2/dt \sim \eta \dot \epsilon^2 \Omega
\end{equation}
The above assumptions lead once again to the Frenkel scaling.

However, this argument appears to be incorrect.
The smoothing out of the cusp at the contact rim is driven by the reduction in interfacial area 
that occurs when the radius $m$ of the meniscus at the edge of the contact area is made larger.
The flow associated with this decrease in the curvature of the meniscus is localized along the contact rim.
Indeed, the flow looks like a line sink at the contact rim; 
the extra volume required to ``fill in'' the neck is drawn from distant regions.
In cross section, this flow must look approximately like a two-dimensional point sink at the contact rim,
with a purely radial velocity (viewed from a point on the rim) varying inversely as the distance to the rim.

The dissipation volume associated with this flow should scale as 
\begin{equation}
\Omega' \sim a m^2
\end{equation}
In the limit in which the meniscus is sharp compared to the contact area (i.e., $m<<a$),
the contact edge may be thought of essentially as a line of length $2 \pi a$,
and clearly then the contact flow cannot ``know about'' the length of the line.

So in that limit, the associated strain rate must be 
\begin{equation}
\dot \epsilon \sim \dot m/m
\end{equation}
The driving force is from reduction in interfacial area and hence energy as the cusp smoothes out,  
\begin{equation}
W' \sim \gamma a m
\end{equation}
again proportional to the length of the edge.

Balancing the dissipation against the rate of change of the interfacial energy
for the secondary neck flow leads to 
\begin{equation}
\dot m \sim \gamma/\eta
\end{equation}

We might have guessed this result on dimensional grounds, 
since it is evident that neither $R$ nor $a$ can enter into the growth rate for $m$
as long as $m$ is much smaller than either of these
(because of the argument that the flow is occurring along a line,
i.e., the flow would look the same in the limit of infinite $a$ and $R$).
Effectively, this secondary flow is equivalent to hydrodynamic coarsening of demixing fluids, 
as described by Siggia \cite{SIGGIA:1979p10393}.

Thus we have $m(t) \sim \gamma t/\eta$. 
How does $m(t)$, which measures the sharpness of the neck cusp, compare to the contact radius $a(t)$?
We may write 
\begin{eqnarray}
& m(t) \sim R t/\tau & \nonumber \\
& a(t) \sim R (t/\tau)^{1/3} & \nonumber \\
& \tau \sim \eta R/ \gamma &
\end{eqnarray}
Here $\tau$ is the timescale for sintering the two drops completely.
We are interested in early times, so $t/\tau<<1$.
Thus it is clear that $m(t) <<a(t)$, so our estimates are consistent ---
the smoothing of the crease at the neck of the sintering droplets
by a secondary flow along the perimeter of the contact area
is a small perturbation on the main scaling of the contact area growth with time.

With a bit more work, we can make a reasonable quantitative estimate 
of the growth rate of the neck radius $a$ due to this secondary flow,
and thereby estimate the value of $a/R$ 
at which the growth rates from the two mechanisms become equal.

\begin{figure}
\begin{center}
\includegraphics[scale=1.7]{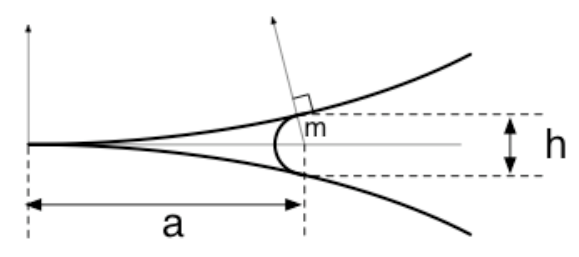}
\caption{\label{neck}Geometry of neck region.}
\end{center}
\end{figure}

We model the geometry of the neck region as a portion of the surface of a torus,
with $a$ the radius of the hole, and $m<<a$ the curvature radius in the orthogonal direction.
(See Fig. \ref{neck}.)  
The gap between the two spherical droplets $h$ satisfies $a^2 = Rh$;
because the droplet radius $R$ is large compared to $a$, 
the curvature radius $m$ is well approximated by $m \approx h/2$.

As the neck advances, fluid must flow in to fill the gap.
When the neck has advanced to a radius $a$,
the volume filled in is 
\begin{equation}
V= 2 \pi \int_0^a \rho d \rho \, h(\rho) = \frac{\pi a^4}{2R}
\label{filled_in}
\end{equation}
This flow is driven by the decrease in exposed surface area of the droplets;
the main contribution is the removal of the nearly-parallel surfaces in the gap,
which have area 
\begin{equation}
A = 2 \pi a^2
\label{area_lost}
\end{equation}
(the factor of 2 comes because we have two droplet surfaces).

The flow towards the gap from elsewhere in the droplets,
we model approximately as a two-dimensional line sink
(bent into a circle of radius $a$).
In cross-section, the two-dimensional incompressible flow into the line sink 
has a velocity $v \approx v_0 m \hat \rho/\rho$,
where $\rho$ is the radial distance from the line sink,
$\hat \rho$ the unit vector pointing away from the line sink,
and $v_0$ some characteristic velocity.
The volume flux to the line sink is
\begin{equation}
\dot V = (2 \pi a) (2 \pi \rho)(v_0 m/\rho) = 2 \pi^2 a^3 v_0/R
\end{equation}
The dissipation from this flow can be estimated as
\begin{equation}
\dot E = (2 \pi a) \int_0^{2 \pi} d \theta \, \int_r^\infty \rho d \rho\, (1/2) \eta (\partial _k v_i + \partial_i v_k)^2 
= 8 \pi^2 a \eta v_0^2
\end{equation}
(In this estimate, we have assumed the flow to the line sink comes from all directions,
and thus neglected the perturbation that comes from the narrow gap.)

Now we determine $v_0$ and the neck growth rate,
by equating the volume flux $\dot V$ to the time derivative of the volume $V$ to be filled in [Eqn.\ (\ref{filled_in})],
and by equating the dissipation rate $\dot E$ to $\gamma \dot A$ from Eqn.\ (\ref{area_lost}).
After a bit of algebra we find
\begin{equation}
\dot a = \frac{\pi \gamma}{2 \eta}
\end{equation}
We can then find the value of $a/R$ at which this estimate of the neck growth rate
equals that from droplet flattening, Eqn.\ (\ref{spreading_result}); 
the two growth rate estimates are equal when $a/R = 1/4$.
This provides an estimate of when we should expect crossover 
between the droplet flattening and cusp smoothing mechanisms for growth of the contact area, 
for sintering of Newtonian fluid droplets.

\section{\label{spreading_expt}Spreading experiment.}
\begin{figure}
\begin{center}
\includegraphics[scale=0.75]{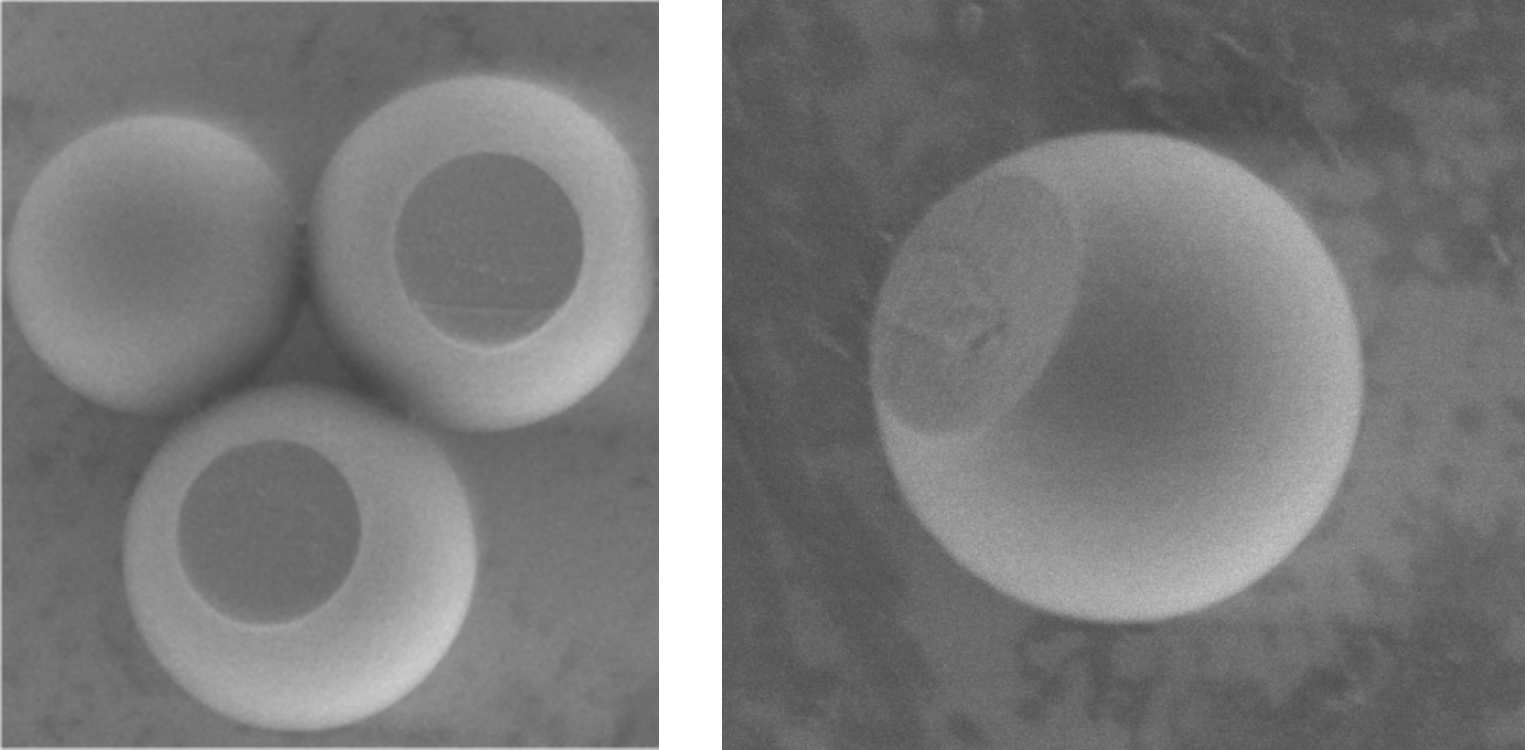}
\caption{\label{Laura_pic}SEM image of $3.3\mu$m diameter PS spheres 
after an interval of spreading at 120C on a silicon substrate.}
\end{center}
\end{figure}

Elegant experiments to investigate early-time droplet spreading  
were performed by Ramirez et al.\ \cite{Ramirez2010}
using 3.3$\mu$m diameter spheres of polystyrene (PS), molecular weight 70kg/mol.
The spheres were settled onto a silicon substrate, heated to 120C
(above their glass transition temperature of $T_g \approx 100$C),
and allowed to spread for a certain time before being quenched.
The spheres were then removed from the substrate
and imaged with a scanning electron microscope (SEM) 
to determine the contact radius $a(t)$ as a function of spreading time $t$.
(See Fig.\ \ref{Laura_pic}.)
Spreading times varied between 20 minutes and 15 hours;
thus at 120C, spreading is slow enough that equilibrating the sample temperature 
is essentially instantaneous.
(Note that the droplet shape at the contact rim is quite sharp,
as we certainly expect for the case of droplet spreading.)

The Newtonian viscosity of the sample was not measured directly,
but was estimated to be about 100MPa.sec.
This estimate was made using literature data for temperature-dependent viscosity of PS,
the sample molecular weight and known molecular weight scaling for entangled polymers, 
and the known temperature dependence of viscosity near the glass transition.
From the known plateau modulus of PS (0.2MPa), 
a terminal time $\tau_\infty$ of about 500sec was estimated.
This is much shorter than the spreading times of the experiments,
thus we expect the spreading PS droplet to behave as a Newtonian fluid.

\begin{figure}
\begin{center}
\includegraphics[scale=0.8]{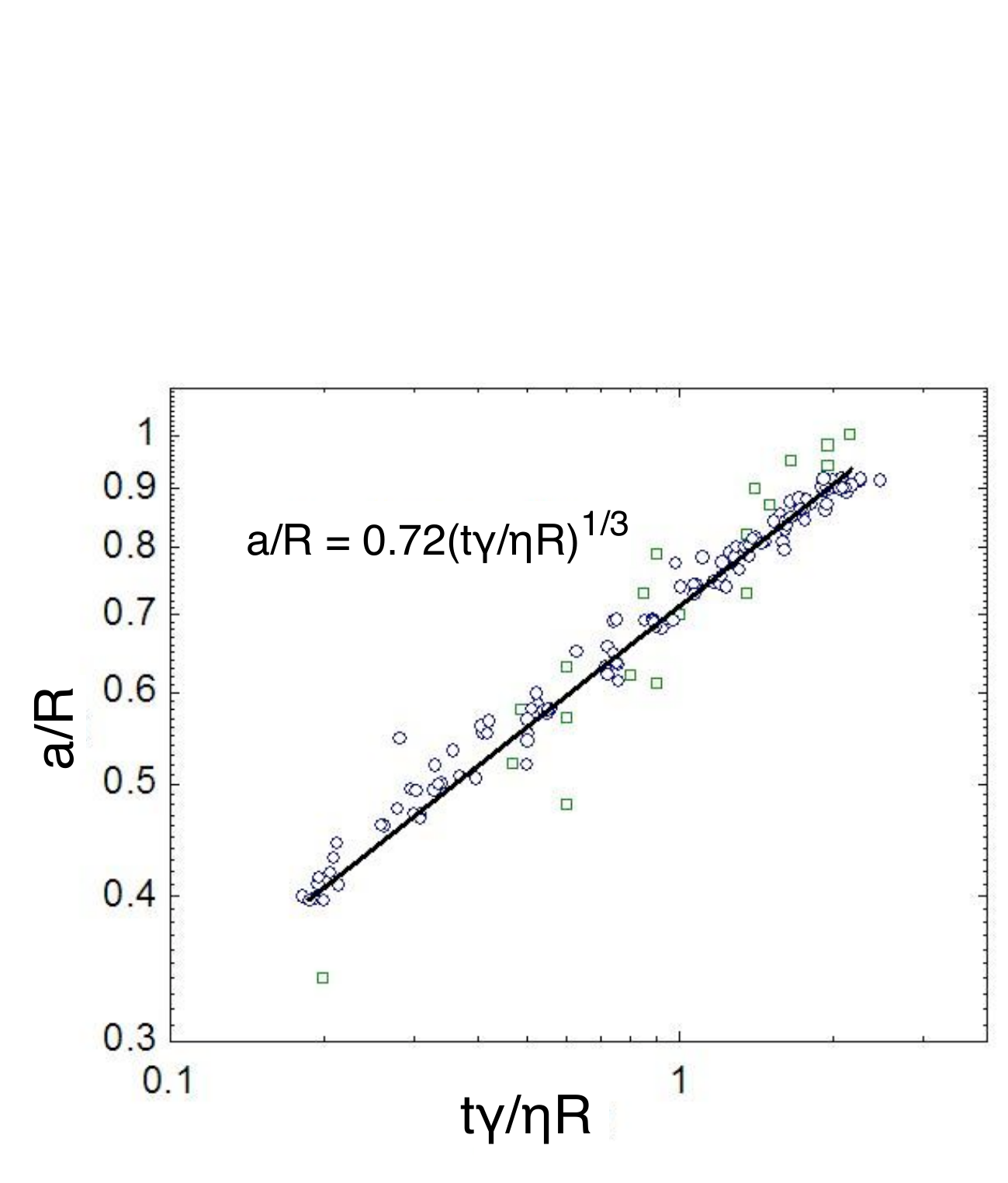}
\caption{\label{Laura_data}Log-log plot of contact radius $a(t)/R$ versus $t/\tau$; solid line is Eqn.\ (\ref{Laura_fit}).
Circles, data of \cite{Ramirez2010}; squares, data of \cite{Bellehumeur:1996p8720}.}
\end{center}
\end{figure}

The measured contact radii $a(t)$ are well described by the expression
\begin{equation}
a(t) = 0.72(t/\tau)^{1/3}
\label{Laura_fit}
\end{equation}
which is the same as Eqn.\ (\ref{contact_vs_time}), 
multiplied by an additional factor of 1.12.  (See Fig.\ \ref{Laura_data}.)

It turns out that the PS droplets do not wet the silicon substrate,
but instead exhibit a rather large contact angle.
However, the PS droplet contact radii follow the $t^{1/3}$ power law uniformly well 
until the contact radius has nearly reached its final value, of about $a=0.9R$.
From the assumption that the droplet shape is a truncated sphere of somewhat larger radius $R'$,
with the same volume as the initial droplet of known radius $R$,
we can infer that the radius of the final truncated sphere is $R'=1.06R$.
The contact angle is then $\theta = \pi - \sin^{-1} a/R'$, or about 122 degrees.

The present theory, appropriate for early time spreading,
assumes a driving force for spreading that is independent of contact angle.
This is fine when the time-dependent contact angle is nearly 180 degrees,
so that the relevant combination of interfacial tensions is 
$\Delta \gamma = \gamma{SV}+\gamma_{LV} - \gamma_{SL}$.
Eventually, for wetting droplets, the contact angle continues to decrease and the droplet thins,
until the geometry of the Tanner law is eventually obtained.
For nonwetting droplets, the Young condition Eqn.\ (\ref{Young_eqn}) is eventually reached,
and the droplet ceases to spread.
Before the final contact radius is reached, the driving force for spreading must steadily decrease to zero,
but as long as the contact radius is well away from the final value, this effect can be neglected.

\section{\label{viscoelastic}Non-Newtonian spreading.}
For polymeric fluids only slightly above their glass transition,
we should expect the spreading or sintering flow to be significantly non-Newtonian.
That is, the stresses in the droplet will have a significant elastic component.

It turns out that we can compute the time-dependent spreading flow
for arbitrary linear viscoelastic response.
We begin by considering two viscoelastic spheres already brought into contact,
with some displacement $h$ and corresponding contact radius $a$,
and then subjected to a small oscillatory motion $\delta h(t) = \Delta h \cos(\omega t)$
of the displacement at some frequency $\omega$.

If we Fourier transform in time, then at frequency $\omega$, 
we have a problem mathematically identical to the previous spreading problem,
since the small displacement $\Delta h$ is spatially uniform over the contact.
We can write
\begin{equation}
v_z(\omega) = \frac{-i \omega J(\omega)}{4 \pi} \int \frac{d^2 r'}{|r-r'|} f_z(r')
\end{equation}

In the above, we have introduced the frequency-dependent compliance $J(\omega)$,
defined as usual:
\begin{eqnarray*}
&\epsilon(\omega) = J(\omega) \sigma(\omega)& \\
&J(\omega) = 1/G(\omega)&
\end{eqnarray*}
The simplest viscoelastic model for $G(\omega)$ is the Maxwell model,
under which we have
\begin{equation}
G(\omega) = \frac{-i \omega G \tau}{1 - i \omega \tau}
\end{equation}

We note that we can write $-i \omega J(\omega) = 1/\eta(\omega)$,
where $\eta(\omega)$ is the complex frequency-dependent viscosity.
(Equivalently, $\eta(\omega)$ is the Fourier transform
of the time-dependent stress relaxation function $G(t)$.)
For the Maxwell model we have $\eta(\omega) = G\tau/(1-i \omega \tau)$.
So the coefficient $-i \omega J(\omega)$ 
in the frequency-dependent velocity equation above,
is just $1/\eta(\omega)$, the frequency-dependent generalization 
of our earlier expression for Newtonian sintering.

Now suppose we impose in the time domain that the velocity $v_z$ 
corresponds to a small displacement step taken at time $t_0$, 
that is, $v_z(t) = (1/2) \Delta h \delta(t-t_0)$.
Note that in specifying this, we are departing (temporarily)
from requiring that the total force be some specified value;
we suppose for the moment that we can apply body forces
to drive the droplets together as we wish.

In the frequency domain, this corresponds to taking 
$v_z(\omega) = (1/2) \Delta h \exp(i \omega t_0)$, 
but spatially constant inside the contact area.
Thus we may apply our previous solution for the force density,
to obtain
\begin{equation}
f_z(r, \omega) = \frac{2 \eta(\omega) \exp(i \omega t_0) \, \Delta h }{\pi a} \frac{1}{\sqrt{1-r^2/a^2}}
\end{equation}
Transforming back to the time domain gives us
\begin{equation}
f_z(r,t) = G(t-t_0) \frac{2 \Delta h}{\pi a} \frac{1}{\sqrt{1-r^2/a^2}}
\end{equation}
where $G(t)$ is understood to be causal, i.e., to vanish for negative argument.

Thus if we drive the displacement with a small step, 
the force density necessary to hold this step 
has the same spatial dependence as for the Newtonian sintering problem,
but a ``fading memory'' given by the usual stress relaxation function.

Now we consider building up a finite displacement history of a growing contact
from a succession of small steps.
(Again we are specifying the displacement history, and computing the required force,
but we will eventually invert that relationship.)
So we write the velocity field as a sum of steps,
\begin{equation}
v_z(t) = (1/2) \sum_i \Delta h_i \delta(t-t_i)
\end{equation}

By superposition, the force areal density is a sum of the responses to these steps,
in which we note that the contact area (over which the force is nonzero)
is of course growing with time.
We are interested in the total force as a function of time,
so as before we integrate $f_z(r,t)$ with respect to area.
For a single step we find
\begin{equation}
F(t) = 4 G(t-t_0) \Delta h \, a(t)
\end{equation}

For a succession of steps, we have
\begin{equation}
F(t) = 4 \sum_i \Delta h_i  G(t-t_i) a(t_i) 
\end{equation}
Taking the limit of small steps as an integral, we have
\begin{equation}
F(t) = 4 \int_{-\infty}^t dt'  \dot h(t') a(t') G(t-t')
\end{equation}

Making use of the relation $2a^2 = Rh$, we see that $\dot h = 4 a \dot a/R$,
and we identify the combination $a^2(t) \dot a(t)$ as $(1/3) d/dt a^3(t)$
so that 
\begin{equation}
F(t) = \frac{16}{3R}\int_{-\infty}^t dt'  \left( \frac{d a^3(t')}{dt'} \right) G(t-t')
\label{creep}
\end{equation}
So the time-dependent force is a convolution 
of the quantity $\dot y(t)$, where $y(t) \equiv a^3(t)$, with the relaxation function $G(t)$.

This relation can be inverted in the frequency domain,
so we Fourier transform to obtain
\begin{equation}
y(\omega) = \frac{3R}{16}J(\omega) F(\omega)
\end{equation}
in which we have used the fact that the Fourier transform of $G(t)$ 
is $\eta(\omega)$ ({\it not} $G(\omega)$, in common rheological notation),
and that $-i \omega J(\omega) = 1/\eta(\omega)$.

We note that $y(\omega) \propto J(\omega) F(\omega)$ 
looks just like the relation $\epsilon(\omega) = J(\omega) \sigma(\omega)$.
Thus the quantity $a^3(t)$ (with its proportionality constant $3R/16$)
responds to the total force $F(t)$ with the usual creep compliance. 
That is, we find 
\begin{equation}
a(t) = \left( \frac{3 \pi \gamma R^2 J(t)}{32} \right)^{1/3}
\label{creep_compliance}
\end{equation}
with precisely the same coefficient as for the Newtonian case,
even though the flow history of different fluid elements are not identical.
If we make the replacement $J(t) \to t/\eta$ appropriate for a Newtonian fluid,
we recover Eqn.\ (\ref{contact_vs_time}).

In the case of sintering, $F(t)$ is a step function at time zero 
(when the particles first start to sinter).
Thus the response of $y(t)$ corresponds to creep after a step stress. 

For well-entangled polymers, we have a well-defined plateau in $J(t)$,
corresponding essentially to deformation under adhesive load of an elastic incompressible solid
with shear modulus equal to the plateau modulus $G_0$.
So we may expect qualitatively to see $a(t)$ quickly reach an apparently steady value
corresponding to a Hertzian contact solution, scaling as $a \sim (\gamma R^2 /G_0)^{1/3}$,
before continuing to grow (on a timescale of order the stress relaxation time of the fluid).

Finally, consider the special case of a power-law stress relaxation function $G(t)$, of the form $t^{-\beta}$. 
Such a stress relaxation function arise when there is such a broad range of relaxation timescales
that the stress relaxation process is approximately timescale invariant.
Typical examples include various multiply-branched polymer melts
(entangled or unentangled critical gels, fractally branched polymers),
with $\beta$ somewhere between zero and unity, and typically near $1/2$.
Then, Eqn.\ (\ref{creep}) implies $d a^3(t)/dt$ is proportional to $t^{\beta-1}$ for positive times,
\begin{equation}
\frac{d a^3(t)}{dt} \propto t^{\beta-1}
\end{equation}
(This form ensures that the right-hand side of Eqn.\ (\ref{creep}) 
is a homogeneous function of $t$ of degree zero, hence a constant for all positive $t$.)
The radius for such a case grows as $t^{\beta/3}$.

Thus non-Newtonian rheology can give rise to growth laws 
for the contact radius different from $t^{1/3}$ at early times,
but typically with {\it faster} growth than $t^{1/3}$ (i.e., $0 < \beta < 1$),
arising ultimately from the contributions of the fastest relaxations
in a broad spectrum of stress relaxation processes.

\section{\label{previous_work}Comparison to prior sintering studies.}
In this section, we briefly summarize the relevant literature on sintering,
divided into analytical, numerical, and experimental approaches for nominally Newtonian fluids,
followed by a discussion of work on sintering of viscoelastic fluids.
This rather extensive literature is somewhat contradictory, as we shall see.

\subsection*{Analytical results.}
A key result was supplied by Hopper \cite{HOPPER:1990p9469}, who used conformal methods 
to construct an exact solution in two dimensions 
for sintering of two parallel cylinders of identical radius,
initially in contact at a single point.
Hopper's solution predicts that at early times, the contact width $w$  
grows according to $w/R \sim (t/\tau) \log (t/\tau)$,
where $\tau=\eta R/\gamma$ is the characteristic time defined previously.
(Of course, our scaling results were constructed from the start 
for the geometry of a circular contact between spherical particles in three dimensions,
so our predictions do not apply to the problem considered by Hopper.)

In later work, Eggers, Lister, and Stone \cite*{Eggers:1999p10625} 
argued that the dominant mechanism for sintering at early times
was the expansion of the ``neck'' between the droplets, driven by capillary forces
resulting from the sharp curvature at the neck.
They asserted that the scaling of this growth mechanism 
should be the same in three dimensions as in two, 
leading to a linear growth law for the contact radius $a/R \sim t/\tau$, with logarithmic corrections.

Our scaling argument for the neck growth rate due to capillary forces 
(see Section \ref{spreading_scaling}) is consistent with the Eggers result.  
However, we assert that the dominant sintering mechanism at early times
is flow in the {\it normal} direction, identical to the early-time spreading flow.
We have shown this flow leads to a growth law $a/R \sim (t/\tau)^{1/3}$,
which should dominate at early times $t << \tau$ and hence $a<<R$.

\subsection*{Numerical approaches.}
The geometry of two spheres (or cylinders, in two dimensions) at the early stages of sintering
is extremely challenging for numerical solutions,
because of the disparity in lengthscales between the contact region and the droplet radius,
or equivalently because of the very sharp curvature of the interface in the contact region.
Early results were limited to two-dimensional problems, 
either using boundary element methods of Kuiken \citep{KUIKEN:1990p10133} 
or finite element approaches of Ross et al.\ \citep{ROSS:1981p10134}.

More recent results of Jagota et al.\ \cite{JAGOTA:1998p10621} and Lin et al.\ \cite{Lin:2001p10619}
presented axisymmetric finite element calculations for a pair of sintering spheres of equal radius.
Log-log plots of the contact radius $a/R$ versus scale time $t/\tau$ 
shows evidence of a $t^{1/3}$ regime at early times, up to perhaps $a/R \approx 0.05$,
before the apparent power-law increases (and then the growth saturates, as the particles coalesce completely).
Corresponding profiles of the interface at different times show a quite sharp cusp until $a/R=0.3$ or so.

Jagota makes a conceptual distinction between a ``zipping'' mode of sintering due to ``surface attraction''
and a ``curvature driven'' neck growth.
In the present work, we would say both of these are driven by reduction in surface free energy,
but differ in the nature of the flow, the first apparently corresponding to the early-time spreading flow,
the second to the secondary flow in the neck region driven by curvature along the neck perimeter.

Most recently, Kirchof, Schmid, and Peukert \cite*{Kirchhof:2009p10622} 
present solutions of the Navier-Stokes equations
including motion of the free surface by the ``fractional volume of fluid'' method.
Their results are consistent with experimental sintering studies of 
Rozenzweig and Narkis \cite{ROSENZWEIG:1981p10141}, 
Bellehumeur et al.\ \cite{Bellehumeur:1996p8720}
and Kingery and Berg \cite{KINGERY:1955p10143}
described below.
Puzzlingly, their numerical results appear to be well-described by the Frenkel scaling,
$a/R \sim (t/\tau)^{1/2}$, without any evident crossover from $t^{1/3}$ to $t$,
as one might expect either on the basis of the present work, or the Jagota work.

\subsection*{Sintering experiments.}
There is a large literature on sintering experiments,
but fewer papers that present careful experimental studies 
on particles with well-controlled spherical geometries, 
made of materials with known rheology,
studied over a wide range of sintering times to obtain wide range of contact radii.

Most studies of arguably Newtonian sintering droplets give results for the contact radius versus time
that are reasonably consistent with the Frenkel scaling 
(which partly explains its longevity as a theory, despite its dubious underpinnings).
Kuczynski \cite{KUCZYNSKI:1949p10149} 
studied small (500 micron) glass spheres, sintered to glass plates;
Kingery and Berg \cite{KINGERY:1955p10143} 
studied somewhat smaller (50 micron) glass spheres in a similar experiment.
Both found time dependence for $a/R$ consistent with Frenkel scaling,
although the range of these data is rather narrow (0.1--0.4 or so) and the points are few.

Later, Rozenzweig and Narkis \cite{ROSENZWEIG:1981p10141} 
studied sintering of pairs of spherical particles of PS or PMMA,
of radii around 500 microns; with $a/R$ ranging from 0.2 to 1, 
they found slopes on a log-log plot versus time between 0.49 and 0.63,
again roughly consistent with Frenkel scaling.
Because the expected numerical prefactor of the Frenkel scaling is not well defined
(as indeed the scaling itself is not theoretically defensible),
it is not possible to check whether the prefactor is consistent 
with the measured Newtonian viscosity of the samples.

Hornsby and Maxwell \cite{HORNSBY:1992p10135} 
studied sintering of pairs of 750 micron diameter spherical particles 
made from various high molecular weight commercial polyproplyenes, at temperatures 180C, 190C, and 200C.
These polymers are viscous not by virtue of proximity to their glass transition,
but simply because the molecular weight is large.
However, the shortest sintering time for which measurements were made (1min)
is much longer than the stress relaxation time in these materials, 
so they may be regarded as Newtonian fluids for the experiment.
Once again, the $a/R$ data in the range 0.2--1 is reasonably consistent with Frenkel scaling, 
with no sign of crossover.

Finally, Yao et al.\ \cite{Yao:2005p10624} set out explicitly
to test the predictions of Eggers et al.\ \cite{Eggers:1999p10625} 
regarding the linear (with log corrections) scaling of neck size with time.
They used a ``Plateau tank'' apparatus, in which two pendant drops attached to constant-pressure reservoirs
face each other, and coalesce when brought into contact.
(Note however that the connection of each pendant drop to its reservoir prevents complete coalescence.)
Yao et al.\ used silicone oils of viscosities ranging from $10^3$ to $10^5$ cSt,
and observed essentially linear initial growth of the neck radius, 
consistent with Eggers et al.\ \cite{Eggers:1999p10625}.

In summary, there is no clear experimental consensus as to the early-time scaling of contact radius 
for sintering droplets of Newtonian fluid.  Most data is roughly consistent with Frenkel scaling,
with the exception of the Yao et al.\ \cite{Yao:2005p10624} experiments, which support a linear growth law.
But most of the data is taken on samples with sparse rheological characterization,
and on rather narrow ranges of $a/R$.
And, in many cases the shortest observation times are uncomfortably close
to typical times to bring the particles and/or the substrate to a steady temperature.

Finally, the Yao setup does not permit the droplets to translate freely towards each other,
which may play a crucial role in shutting down what we would otherwise predict to be the dominant flow,
in which the particles progressively move towards each other and flatten, to quickly increase the contact area.

\subsection*{Non-Newtonian sintering.}
Several previous authors have attempted to analyze sintering of non-Newtonian fluids,
sometimes with some time-dependent theory ultimately based on Frenkel scaling
\cite{Bellehumeur:1998p10620},
and more recently \cite{Scribben:2006p10616} with an elaborate approach 
based on crack healing in viscoelastic materials based on work of Schapery \cite{SCHAPERY:1989p10630}
and Hui et al.\  \cite{Hui:1998p10633}.

For the present work, the most significant prior results in this area 
are those of Mazur and Plazek \cite{MAZUR:1994p9603},
who anticipate the main result of Section \ref{viscoelastic} above.
They make the heuristic argument that sintering of viscoelastic particles on a timescale $t$
is equivalent to formation of a Hertzian contact under adhesive load
of particles with an effective compliance $J(t)$.
As shown in Section \ref{viscoelastic}, this argument turns out to be precisely correct,
even to the numerical prefactor.

Mazur and Plazek \cite{MAZUR:1994p9603} 
emphasize the variations in the apparent power law relation
between $a/R$ and $t$, ranging from 0.2 to 0.9,
for sintering studies of materials that are arguably not Newtonian 
(though often the rheology in such studies is ill-characterized).
Such behavior, they argue, can arise from viscoelastic sintering.
They performed sintering experiments by observing with an optical microscope 
the spreading of a 250 micron diameter polymer droplet on a thick polymer substrate, slightly above $T_g$.
The material studied was a poly methyl methacrylate / ethyl methacrylate random copolymer (PMMA/PEMA),
with a measured creep compliance varying over five orders of magnitude from 1 to $10^5$ sec.

\begin{figure}
\begin{center}
\includegraphics[scale=0.6]{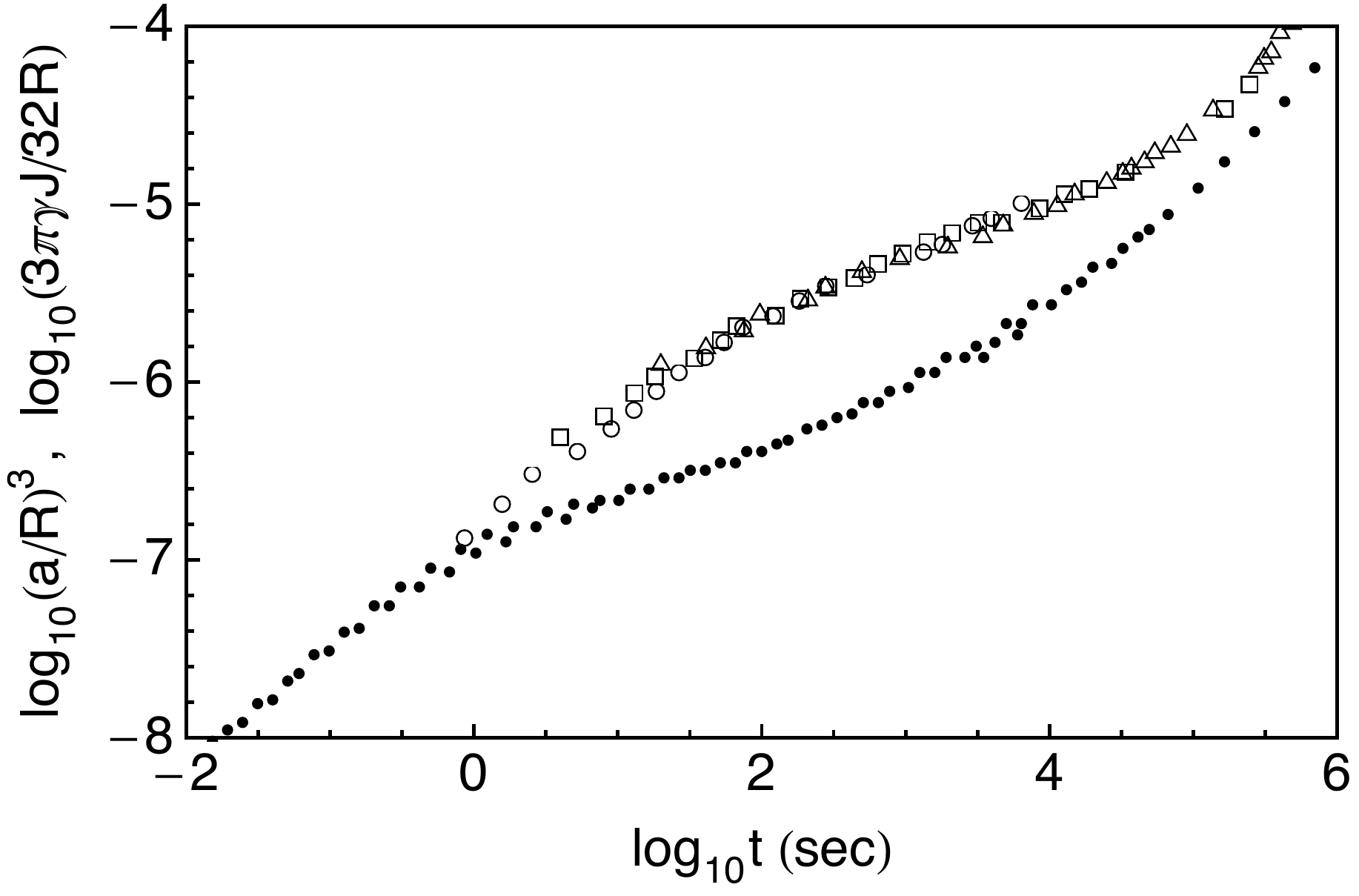}
\caption{\label{plazek_fig}Data of \cite{MAZUR:1994p9603}.  
Compliance $J(t)$ (small points) and sintering radius plotted as $(a(t)/R)^3$ 
(open symbols, data at 120C, 132C, 140C), log-log versus time $t$, master curve at 132C.
Sintering data vertically shifted.}
\end{center}
\end{figure}

In Fig.\ \ref{plazek_fig}, we plot the results of Mazur and Plazek \cite{MAZUR:1994p9603}
as $(a(t)/R)^3$ and $3 \pi \gamma J(t)/(32 R)$ versus time (log-log),
to test the validity of our result Eqn.\ (\ref{creep_compliance}) 
for the spreading and sintering of non-Newtonian fluids.
The neck radius and compliance data have both been time-temperature shifted 
to produce a master curve at 132C.

Qualitatively, the shapes of $a(t)/R^3$ and the predicted values 
$3 \pi \gamma J(t)/(32 R)$ are suggestively similar.
(Surprisingly, after having made the argument that $(a(t)/R)^3$ should be proportional to $J(t)$,
Mazur and Plazek did not actually plot the data this way, but only compared the shapes of $a(t)/R$ and $J(t)$.)

The estimated crossover point between droplet flattening and neck growth mechanisms, 
$a/R \doteq 0.14$, occurs at a time of about 60 sec.
Significant discrepancies between the predicted and measured neck radii
evidently appear in the data of Fig.\ \ref{plazek_fig} at somewhat earlier times.
Recall however that the crossover estimate was made for Newtonian fluids only.
  
Mazur and Plazek \cite{MAZUR:1994p9603} emphasize the likely prevalence of viscoelastic sintering 
in practical situations, for submicron-sized particles such as are present for example in a deposited latex film.
As the radius $R$ of the particles decreases, the characteristic time $\tau=\eta R/\gamma$ 
for them to sinter by viscous flow likewise decreases,
until it becomes comparable to some characteristic viscoelastic relaxation time $\tau_0$ of the material.
The size $R^*$ at which $\tau = \tau_0$ is $R^* = \gamma \tau_0/\eta \approx \gamma/G$,
where $G$ is the plateau modulus and we have approximated $\eta \approx G \tau_0$.
As a concrete example, consider polyisobutylene, with a plateau modulus $G \approx 0.2$MPa,
an interfacial tension $\gamma$ of perhaps 20 mN/m.
Thus $R^* \approx 0.1 \mu$m; particles smaller than 0.1$\mu$m will sinter viscoelastically.

\section{Conclusions}
We have constructed a scaling theory and a detailed calculation
of the dynamics of spreading of a droplet on a solid substrate at early times,
when the contact radius $a$ is small compared to the droplet radius $R$,
by analogy to the displacement field of a Hertzian elastic contact.
The flow within the droplet progressively flattens the contact area,
and is closely related to the displacement field of the flattened elastic contact.

In the scaling description, balancing the rate of viscous dissipation
with the rate at which the interfacial energy decreases as the droplet spreads
results in the contact radius scaling as $a(t)/R \sim (t/\tau)^{1/3}$,
where $\tau = \eta R/\gamma$ is the characteristic spreading time.
The detailed calculation gives $a(t)/R=(3 \pi \gamma t/(32 \eta R))^{1/3}$,
consistent with the scaling result.
Our result is complementary to the well-known Tanner law for droplet spreading, $a(t)/R \sim (t/\tau)^{1/10}$,
which applies when the droplet has spread far beyond its initial radius 
and the contact angle is therefore small.

We have argued that our description of droplet spreading at early times
also applies to the early-time dynamics of droplet sintering.
The hydrodynamics of the two problems are not identical,
because the boundary conditions on the contact plane are different 
(zero velocity for spreading, zero transverse stress for sintering).

In the sintering problem, we can have radial flows in the contact plane
driven by the transverse curvature of the ``neck'' region between the droplets.
However, we have given scaling arguments showing that the transverse curvature radius $r$
grows with time as $r(t) \sim \gamma t /\eta$, driven by smoothing of the cusp at the neck. 
Thus $r(t)/R \sim (t/\tau)$, so that for early times $r$ is small, the cusp is sharp, 
and the contact radius is determined by the same flows
as describe the early dynamics of droplet spreading.
We construct a model calculation for the growth rate of the neck by cusp smoothing,
and find a crossover value of $a/R \doteq 0.14$ between droplet flattening and cusp smoothing mechanisms
for growth of the contact radius.

Our results contradict the scaling arguments of Frenkel \cite{Frenkel},
who long ago proposed that the contact radius of particles sintering by viscous flow
should grow as $a(t)/R \sim (t/\tau)^{1/2}$.
We show that Frenkel's approach, as well as later attempts to redeem his result,
make incorrect assumptions as to the length scales for flow gradients and the dissipation volume.

More recent experiments on droplet spreading at early times 
were performed by Ramirez et al.\ \cite{Ramirez2010},
using micron-sized polystyrene (PS) spheres of molecular weight 70kg/mol and a silicon substrate.
Contact radii were determined with scanning electron microscopy after a finite spreading time.
The spreading was conducted at 120C (above the nominal $Tg$ of PS of about 100C);
the stress relaxation time was estimated to be short enough that the spreading flow was Newtonian.
The droplets were nonwetting on silicon, with a final contact angle of about 120 degrees,
but contact radii at earlier times were found to be well described by our results.

The computational and experimental literature for sintering is much less clear.
Numerical results for sintering particles are challenging, 
because of difficulty of representing sharp cusp at joint between particles.
Results for the early-time growth of the contact radius vary considerably.
Classical literature data on sintering particles is generally in agreement with Frenkel scaling
(which is not defensible theoretically), albeit most data cover a relatively narrow range of $a/R$ values,
and were obtained using samples of uncertain rheology.

In commercial practice as in many experiments, sintering droplets are often viscoelastic. 
Data and analysis of Mazur and Plazek \cite{MAZUR:1994p9603} emphasizes the importance 
of accounting for time-dependent compliance in interpreting sintering data,
particularly for smaller particles.

To resolve lingering issues about the appropriate description of sintering flows,
we propose a series of experiments, using the general approach of Mazur and Plazek \cite{MAZUR:1994p9603}
(in-situ microscopy observations of polymeric droplets spreading on identical polymer substrates, 
slightly above $T_g$).
Carefully chosen model polymers (e.g., anionically synthesized polystyrene) 
will have narrow molecular weight distribution, and thus a relatively well-defined viscoelastic timescale.
Operating slightly above $T_g$ slows down the spreading to a convenient rate,
allowing for nearly-instant equilibration of the temperature compared to spreading times.
The thickness of the polymer substrate can be systematically varied,
to pass from spreading (where the cusp-smoothing mechanism is suppressed 
by the no-slip boundary condition on the substrate)
to sintering (when the substrate polymer layer is thick compared to the dimensions of the contact region).

\begin{acknowledgments}
We thank Darrell Velegol and Ralph Colby for useful discussions, 
Laura Ramirez for sharing her data and providing figures prior to publication,
and Joe McDermott for bringing this problem to our attention.
\end{acknowledgments}

\bibliography{sintering}

\end{document}